\documentclass[11pt]{JHEP3} 

\skip\footins = 1\bigskipamount plus 2pt minus 4pt

\usepackage{amsmath}
\usepackage{epsfig}
\usepackage{subfigure}
\usepackage{amssymb}

\newcommand{\eq}[1]{eq.~(\ref{#1})}
\newcommand{\eqs}[1]{eqs.~(\ref{#1})}
\newcommand{\meq}[1]{(\ref{#1})}
\newcommand{\tm}{\tilde M}
\newcommand{\grad}{\nabla}
\newcommand{\pa}{\partial}
\newcommand{\pf}[2]{\frac{\pa #1}{\pa #2}}
\newcommand{\oh}{\frac{1}{2}}

\title{The covariant entropy bound in gravitational collapse}

\author{Sijie Gao  and  Jos\'e P. S. Lemos \\
Centro Multidisciplinar de Astrof\'{\i}sica - CENTRA\\
Departamento de F\'{\i}sica, Instituto Superior T\'ecnico\\
Av.\ Rovisco Pais 1, 1049-001 Lisboa, Portugal\\
E-mail: \email{sijie@fisica.ist.utl.pt}, \email{lemos@fisica.ist.utl.pt}}

\abstract{We study the covariant entropy bound in the context of
  gravitational collapse. First, we discuss critically the heuristic
  arguments advanced by Bousso. Then we solve the problem through an
  exact model: a Tolman-Bondi dust shell collapsing into a
  Schwarzschild black hole. After the collapse, a new black hole with
  a larger mass is formed. The horizon, $L$, of the old black hole
  then terminates at the singularity. We show that the entropy
  crossing $L$ does not exceed a quarter of the area of the old
  horizon. Therefore, the covariant entropy bound is satisfied in this
  process.}

\begin{document}

\section{Introduction}

Bousso's covariant entropy conjecture says~\cite{bousso}: let $A$ be
the area of a connected two-dimensional spatial surface $B$ contained
in a four-dimensional spacetime ${\cal M}$. Let $L$ be a hypersurface
bounded by $B$ and generated by surface orthogonal null geodesics with
non-positive expansion. Then the total entropy, $S$, contained on $L$
can not exceed one quarter of the area, i.e., $S\leq A/4$.

Evidences supporting this conjecture have been found in cosmological
solutions and other matter systems~\cite{bousso}--\cite{bthree}. One
example discussed by Bousso~\cite{bousso} is a shell collapsing into a
black hole. We now briefly review the arguments given
in~\cite{bousso}.  Consider a shell with mass $M$, collapsing into a
black hole with original mass $M_0$. The apparent horizon $L$ is
located at $r=r_0$ before the collapse. Once the shell has crossed the
light-sheet $L$ at $r_0$, the null generators of $L$ will be moving in
a Schwarzschild interior of mass $\tilde M=M+{r_0}/{2}$ and hit
the singularity $r=0$ eventually. Since only the entropy crossing $L$
is relevant in the entropy bound, we consider the case where the outer
surface of the shell meets $L$ right at the singularity (See
figure~\ref{shell}).

\FIGURE[t]{\centerline{\epsfig{file=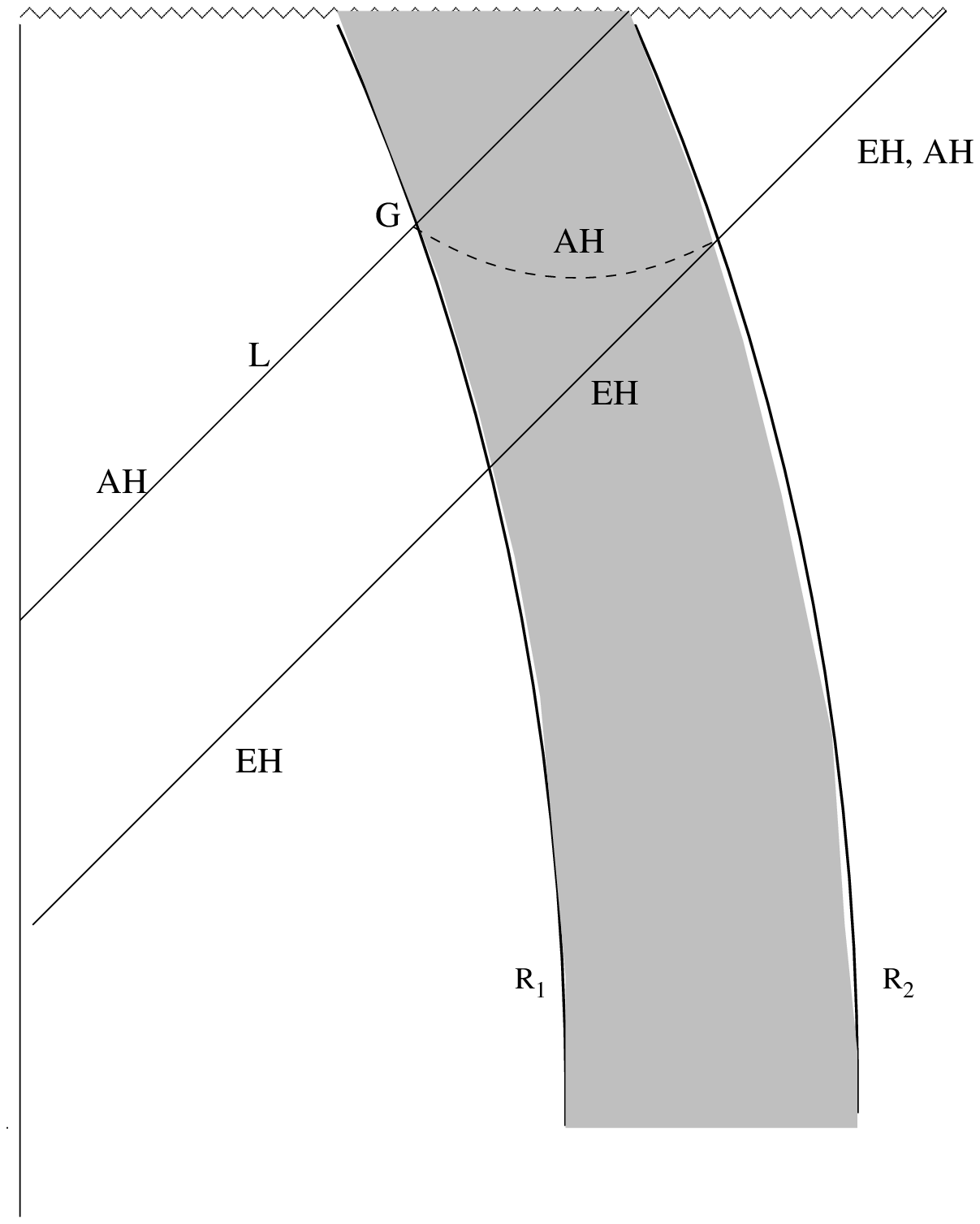, width=.5\textwidth}}%
\caption{A Tolman-Bondi dust shell falling into a black hole. AH
  represents the apparent horizon and EH represents the event
  horizon.\label{shell}}}

The author in~\cite{bousso} claims that the meeting occurs at proper
time
\begin{equation} 
\Delta \tau_{\rm dead}=r_0+2\tilde M \ln \left(1-\frac{r_0}{2\tilde
  M}\right)\approx \frac{r_0^2}{4\tm}\,.
\label{tde} 
\end{equation}
The approximation above was made under the condition $r_0\ll 2\tm$ or
$r_0\ll 2M$.  The maximum proper thickness of the shell is then
approximated as
\begin{equation} 
w_{\rm max}\approx \frac{r_0^2}{2\tm}\,. \label{wm}
\end{equation}
Next, the author used the Bekenstein's bound to estimate the maximum
entropy of shell and obtained
\begin{equation} 
S=2\pi w M\,. \label{ms}
\end{equation}
The condition $w\leq w_{\rm max}$ gives 
\begin{equation} 
S_{\rm max}=\pi r_0^2\frac{M}{\tm}\leq\frac{A}{4}. \label{sma}
\end{equation}
The expression above shows that the covariant entropy bound is
preserved. We see that the maximum thickness of the shell plays an
essential role in deriving the bound. However, the author did not show
explicitly how $\Delta \tau_{\rm dead}$ was derived and did not
explain clearly its physical meaning. We shall show that the
right-hand side of \eq{tde} can only be interpreted as a coordinate
distance. The Schwarzschild metric takes the form
\begin{equation} 
ds^2=-\left(1-\frac{2\tilde M}{r}\right)dt^2+\left(1-\frac{2\tilde
  M}{r}\right)^{-1}dr^2+r^2 d\Omega^2\,. \label{tm}
\end{equation}
Inside the black hole the coordinate $t$ plays the role of distance,
while $r$ plays the role of time.  For a null geodesic starting at
$r=r_0<2M$ and ending up at the singularity $r=0$, the difference in
$t$ is
\begin{eqnarray}
\Delta t &=&\int_0^{r_0}\frac{r}{r-2\tilde M}dr 
\nonumber \\
&=& r_0+2\tilde M \ln \left(1-\frac{r_0}{2\tilde M}\right)\approx
\frac{r_0^2}{4\tilde M}\,, 
\label{delt}
\end{eqnarray}
which is just the right-side of \eq{tde}.  From our calculation above,
the interpretation of $\Delta t$ is clear: inside the black hole,
$r<2\tilde M$, there are comoving observers associated with
coordinates $(t,r)$ with $-({\partial}/{\partial r})^a$
tangent to the world lines of the observers. Different from the
outside of the black hole, these observers are no longer
stationary. Suppose that we choose an observer $A$ and at the moment
$r=r_0$, an outgoing light signal is sent from this observer in the
radial direction and reach another observer $B$ at the singularity
$r=0$. Then, $\Delta t$ is the coordinate distance between $A$ and
$B$. The observer $A$ corresponds to the middle of the collapsing
shell and $B$ to the outer surface. According to~\cite{bousso}, the
finite null geodesic will restrict the thickness of the
shell. Eq.~(\ref{tde}) is used in~\cite{bousso} to estimate the upper
bound of the proper thickness of the shell given in \eq{wm}. However,
the proper distance, $\Delta \tau$, between two comoving observers at
the moment $r_0$ differs from the coordinate distance~by
\begin{equation} 
\Delta \tau=\Delta t\sqrt{\frac{2\tilde M}{r_0}-1}\,. 
\label{prd}
\end{equation}
For $r_0\ll 2\tilde M$, we have $\Delta \tau\gg\Delta t$. This means
that the maximum width of the shell can be much larger than what is
given in~\cite{bousso}.  Now, we shall argue furthermore that the
thickness of the shell may be arbitrarily large, i.e., the idea that a
finite light ray gives rise to an upper bound of proper distance
between observers passing through it is simply wrong. The above
maximum thickness is obtained by assuming that the shell moves at the
moment $r=r_0$ with the comoving observers inside the black
hole. There is no explanation in~\cite{bousso} why this particular
motion is preferred.  Now we show the proper length can be arbitrarily
large if motions of the shell can be arbitrarily chosen. The argument
can be easily illustrated in the case of special relativity.

\FIGURE[t]{\centerline{\epsfig{file=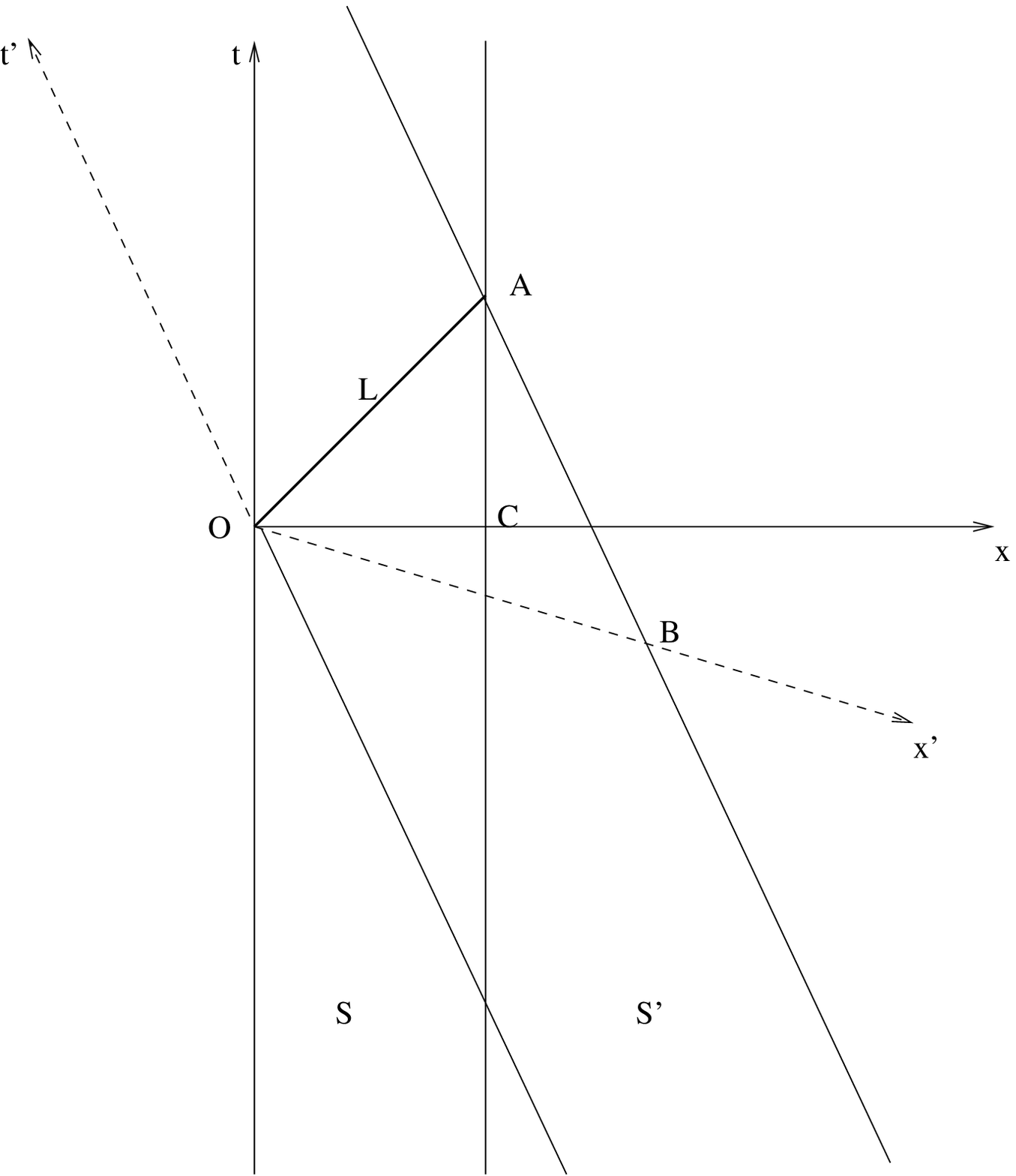, width=.5\textwidth}}%
\caption{A spacetime diagram for observers in references $S$ and $S'$
  passing through the light-sheet $L$ in Minkowski spacetime,
  respectively. The proper distances between the two observers in $S$
  and $S'$ are $|OC|$ and $|OB|$, respectively.\label{shell-mk} }}

In figure~\ref{shell-mk}, $L$ represents a light sheet in Minkowski
spacetime. S and S' are two inertial reference frames with coordinates
$(t,x)$ and $(t',x')$, respectively. Let $v$ be the relative velocity
between S and S'. Then the Lorentz transformation is given by the two
equations:
\begin{eqnarray}
t'&=&\gamma (t-vx)\,, 
\nonumber \\
x'&=&\gamma (x-vt)\,. 
\label{lt}
\end{eqnarray}
Figure~\ref{shell-mk} corresponds to a negative $v$. The proper length
between the two observers in S passing through $L$ is $|OC|\equiv
L_0$. Here $L_0$ is analogous to the proper length of the shell which
follows the comoving observers inside the black hole. It is not
difficult to find that the coordinates of $B$ in S are
\begin{eqnarray}
t_B&=&\frac{v}{1+v}L_0\,, 
\nonumber \\
x_B &=& \frac{L_0}{1+v}\,. 
\end{eqnarray}
Then the proper length of $OB$ is easily obtained from the Minkowski
metric
\begin{equation} 
|OB|=L_0\sqrt{\frac{1-v}{1+v}}\,. 
\label{ob} 
\end{equation}
Therefore, when $v$ is arbitrarily close to $-1$, $|OB|$ will be
arbitrarily large, i.e., the finite light sheet $L$ can not put an
upper bound on the proper length between observers passing~it.

One can generalize the arguments above with some ease to a curved
background. So the estimation of the maximum width of the shell given
in~\cite{bousso} is not valid. Another problem in Bousso's analysis is
that the shell of finite thickness is simplified by an infinitely thin
shell. So the details of the shell become irrelevant. It is unclear to
us whether this approximation is valid or accurate.

To clarify these issues, we first construct an exact solution of a
shell filled with Tolman-Bondi dust collapsing into a black hole. The
regions inside the inner boundary and outside the outer boundary of
the shell are described by Schwarzschild metrics with different
masses.  Following the discussion on the homogeneous cosmology
in~\cite{bousso}, we assume that the entropy of the shell is invariant
in a comoving volume. We then test the bound for $L$ during the
collapse. Our results show that there is no violation for the
covariant entropy bound in the classical regime.

\section{A thick shell collapsing into a black hole}

Relativistic thin shell solutions have been studied by many authors
(see e.g.,~\cite{israel}--\cite{two}). Thin shells are considered as
zero thickness objects. Consequently, the extrinsic curvature across
the shell becomes discontinuous. The jump of the extrinsic curvature
is related to the surface stress-energy tensor. A spherical thick
shell solution was found and corrections to the thin shell dynamics
were discussed by Khakshournia and Mansouri~\cite{thick}. The shell is
immersed in two different spherically symmetric space-times. In
contrast to thin shell cases, the junction conditions for thick shells
then become that the extrinsic curvatures across the inner boundary
and the outer boundary of the shell are continuous, respectively. We
apply these conditions to a spacetime where a spherical thick shell is
immersed in two Schwarzschild spacetimes.  In this solution, a
Schwarzschild black hole with mass $M_0={r_0}/{2}$ is already
there. Then a shell with gravitational mass $M$ collapses into it and
forms a new black hole with mass ${\tilde M}=M_0+M$.  The shell
consists of Tolman-Bondi dust. One assigns to each layer of the thick
shell a comoving coordinate $R$ with an energy per unit mass
$E(R)$. We will only discuss the marginally bound case with $E(R)=0$,
which is described by the metric~\cite{lemos91}
\begin{equation} 
ds^2=-dT^2+r'^2dR^2 +r^2(T,R)(d\theta^2+\sin^2\theta d\varphi^2)\,, \label{tbm}
\end{equation}
where $R_1\leq R \leq R_2$ and $r'={\partial r}/{\partial R}$. We
denote by $\Sigma_1$ and $\Sigma_2$ the inner boundary $R=R_1$ and the
outer boundary $R=R_2$, respectively.  The function $r(T,R)$ is the
solution of
\begin{equation} 
\oh \left(\frac{\partial r}{\partial T}\right)^2=\frac{M(R)}{r}\,,
\label{mer}
\end{equation}
where $M(R)$ is the effective gravitational mass within $R$. The
self-similar solution in this case is~\cite{lemos91}
\begin{eqnarray}
M(R)&=&R \,,
\nonumber\\
r(T,R) &=&\left(\frac{9}{2} R\right)^{1/3}\left(BR-T\right)^{2/3} ,
\label{sss}
\end{eqnarray}
where the constant $B$ is a homogeneity parameter.  The shell divides
the spacetime into three regions. Denote the region inside $\Sigma_1$
by ${\cal M}_{\rm in}$, outside $\Sigma_2$ by ${\cal M}_{\rm out}$ and
the shell by ${\cal M}_{\rm s}$.  ${\cal M}_{\rm in}$ is described by
the Schwarzschild metric:
\begin{equation} 
ds^2=-\left(1-\frac{2M_0}{r}\right)dt^2+\left(1-\frac{2M_0}{r}\right)^{-1}
dr^2+r^2 d\theta^2+r^2 \sin^2\theta d\varphi^2 \,,
\label{min}  
\end{equation}
where $r<r(T,R_1)$. $M_0$ needs to be determined by boundary
conditions of $\Sigma_1$.  $\Sigma_1$ is a timelike hypersurface
described by $R=R_1$. Following~\cite{thick}, we require the
continuity of the extrinsic curvature tensor $K_{ab}$ of
$\Sigma_1$. To calculate $K_{ab}$, consider the tangent vector of a
comoving observer on $\Sigma_1$:
\begin{equation} 
\left(\pf{}{T}\right)^a= \dot t\left(\pf{}{t}\right)^a 
+\dot r \left(\pf{}{r}\right)^a, 
\label{ptm}
\end{equation}
where $\dot r={\pa r(T,R_1)}/{\pa T}$. $\dot t$ can be solved by the
fact that $T$ is the proper time and therefore $g_{ab}({\pa}/{\pa
  T})^a ({\pa}/{\pa T})^b=-1$ in ${\cal M}_{\rm s}$.  The matching
conditions require that $({\pa}/{\pa T})^a$ be also normalized
when measured by the metric of ${\cal M}_{\rm in}$. Hence, from
\eqs{min} and~\meq{ptm}, we have
\begin{equation} 
-\left(1-\frac{2M_0}{r}\right)\dot t^2 +
\left(1-\frac{2M_0}{r}\right)^{-1} \dot r^2 =-1\,, 
\label{norm}
\end{equation}
and then
\begin{equation} 
\dot t= \left(1-\frac{2M_0}{r}\right)^{-1/2}  \sqrt{1+
  \left(1-\frac{2M_0}{r}\right)^{-1} \dot r^2 }\,.
\end{equation}
The unit tangent field orthogonal to $\Sigma_1$ in ${\cal M}_{\rm in}$
can be calculated from \eqs{min} and~\meq{ptm}~as
\begin{equation} 
n_a=-\dot r (dt)_a+
\left(1-\frac{2M_0}{r}\right)^{-1}\sqrt{1-\frac{2M_0}{r}+\dot r^2}(dr)_a\,.  
\label{inn}
\end{equation}
The extrinsic curvature, $K_{ab}$, of $\Sigma_1$ is defined
by~\cite{gr}
\begin{equation} 
K_{ab}=\grad_a n_b\,. \label{kab}
\end{equation}
Straightforward calculation shows that the $\theta$-$\theta$ component
of $K_{ab}$ is
\begin{equation} 
K_{\theta\theta}=r\sqrt{1-\frac{2M_0}{r}+\dot r^2}\,.
\label{kthi}
\end{equation}
It is understood that all quantities are valuated on $\Sigma_1$.
Similar calculation gives the $\theta$-$\theta$ component of $K_{ab}$
derived from the metric of ${\cal M}_{\rm s}$:
\begin{equation} 
K_{\theta\theta}=r\,. \label{ktms}
\end{equation}
By continuity, the right-hand sides of \eqs{kthi} and \meq {ktms} must
be equal. Then, from \eqs{mer} and~\meq{sss}, we get immediately
\begin{equation} 
M_0=R_1\,. 
\label{mro}
\end{equation}
This result is expected since $R_1$ would be the gravitational mass
within $\Sigma_1$ if the region of ${\cal M}_{\rm in}$ were filled
with Tolman-Bondi dust described by \eq{tbm}. Similarly, the spacetime
exterior to the shell is Schwarzschild with mass $\tilde M=R_2$. In
summary, the metric of the entire spacetime is
\begin{equation} 
ds^2{=}\left\{ 
\begin{array}{ll}
\displaystyle  
{-}\bigg(1-\frac{2M_0}{r}\bigg)dt^2
{+}\bigg(1-\frac{2M_0}{r}\bigg)^{-1}
dr^2{+}r^2 d\Omega^2\,,  
& {\cal M}_{\rm in} 
\\[7pt]
\displaystyle  
{-}dT^2{+}\frac{(3 BR-T)^2 }{R^{4/3}(6B\,R-6T)^{2/3}}dR^2
{+}\bigg(\frac{9}{2}R\bigg)^{2/3}(B\,R-T)^{4/3} d\Omega^2\,,  
\qquad & R_1<R<R_2
\\[7pt]
\displaystyle  
{-}\bigg(1-\frac{2\tilde M}{r}\bigg)dt^2
{+}\bigg(1-\frac{2\tilde M}{r}\bigg)^{-1}
dr^2{+}r^2 d\Omega^2\,, 
& {\cal M}_{\rm out} 
\label{thm}
\end{array}\right. 
\end{equation}
where $M_0=R_1$ and  $\tilde M=R_2$. A future singularity is located
at~\cite{lemos91}
\begin{equation} 
T=BR\,. \label{tbr}
\end{equation}

Since both ${\cal M}_{\rm in}$ and ${\cal M}_{\rm out}$ are
Schwarzschild, the apparent horizon satisfies $r=2M_0$ in ${\cal
  M}_{\rm in}$ and $r=2\tilde M$ in ${\cal M}_{\rm out}$. To find out
the apparent horizon inside the shell ${\cal M}_{\rm s}$, we calculate
the expansion, $\theta$, of the outgoing null geodesics.  The apparent
horizon follows the equation $\theta=0$, which yields
\begin{equation} 
T=\bigg(-\frac{4}{3}+B\bigg)\,R  \,,
\qquad 
R_1\leq R\leq R_2 \, .
\label{aps} 
\end{equation}
Now we show that this solution makes the apparent horizon in the
spacetime continuous.\footnote{The apparent horizon was defined by
  Hawking~\cite{hawking} to be the outer boundary of the total trapped
  region in a spacelike slice. In that case, the apparent horizon
  moves discontinuously. In this paper, we adopt the generalized
  definition that the apparent horizon is the boundary of the total
  trapped region~\cite{smarr} (see also~\cite{gr}).} Denote by $G$ the
intersection of $L$ and $\Sigma_1$ (see figure~\ref{shell}). Note that
$G$ is a 2-sphere. By continuity of spacetime, its proper area
calculated from ${\cal M}_{\rm in}$ side and ${\cal M}_{\rm s}$ side
should be equal. Since the radii of 2-spheres along $L$ in ${\cal
  M}_{\rm in}$ are always $2R_1$, we have immediately
\begin{equation} 
2R_1=r(T,R_1)\,, 
\label{met}
\end{equation}
which gives
\begin{equation} 
T=\bigg(-\frac{4}{3}+B\bigg)\,R_1\,.   
\label{wgi} 
\end{equation}
Similarly, we have the meeting time of the apparent horizon in ${\cal
  M}_{\rm out}$ with $\Sigma_2$:
\begin{equation} 
T=\bigg(-\frac{4}{3}+B\bigg)\,R_2\,.  
\label{wgh} 
\end{equation}
Thus, the apparent horizon in ${\cal M}_{\rm s}$ determined by
\eq{aps} joins the other two portions of the apparent horizon in a
continuous manner (see figure~\ref{shell}).

\section{Testing the covariant entropy bound}

In this section, we shall test the covariant entropy bound in the
collapsing process. The light sheet is chosen to be $L$ with area
$A=4\pi (2M_0)^2$. To specify the entropy of the shell, we generalize
the prescription in~\cite{bousso} where entropies for cosmological
models are defined. By applying the coordinate transformation
\begin{equation} 
\bar R =3R^{1/3} 
\label{rtr}
\end{equation}
to the metric~\meq{thm} with $B=0$, we obtain the flat dust-filled
Robertson-Walker universe:
\begin{equation} 
ds^2=-dT^2+a^2(T)(d\bar R^2+\bar R^2 d\Omega^2)\,, 
\label{frf}
\end{equation}
where $a(T)=\frac{1}{6^{1/3}}\,T^{2/3}$. According
to~\cite{bousso,bn}, the universe evolves adiabatically and
consequently, the physical entropy density, $s(T)$, is diluted by
cosmological expansion:
\begin{equation} 
s(T)=\frac{s_0}{a^3(T)}\,. 
\label{pe}
\end{equation}
The constant $s_0$ can be specified by assuming that the entropy may
not exceed one per Plank volume at the Planck time
$T_{Pl}$\footnote{In~\cite{bousso}, the Planck time is near a past
  singularity. In our case, a future singularity is encountered. Thus,
  we shall use time symmetry and impose a future boundary
  condition.}~\cite{bousso}. Therefore, we choose $s$ such that
\begin{equation} 
s(T)=\frac{a^3(T_{Pl}) }{a^3(T)}\,. 
\label{nden}
\end{equation}
To generalize the above treatment to the inhomogeneous case, we make
the following modifications. During the inhomogeneous collapse, the
physical entropy density $s$ is a function of both $T$ and $R$. We
denote it by $s (T,R)$. In analogy to the dust universe, we assume
that the shell evolves adiabatically. Thus, the entropy in a comoving
volume is constant. From the metric~\meq{tbm}, we have the volume,
$dV$, in a thin shell between $R$ and $R+dR$:
\begin{equation} 
dV=4 \pi r'(T,R)r^2(T,R)dR\, . 
\label{dv}
\end{equation}
Thus
\begin{equation} 
s(T,R)dV=4 \pi\, s(T,R) r'(T,R)r^2(T,R)dR  
\label{sdv}
\end{equation}
should be independent of $T$ for fixed $dR$. Therefore,
\begin{equation} 
4\pi\, s(T,R) r'(T,R)r^2(T,R)\equiv S(R)\,, 
\label{ssr}
\end{equation}
where $S(R)$ is the entropy per unit length.  We still require that
the entropy density $s$ be smaller than one at the ``Planck
time''. But the ``Planck time'' here is no longer equal to $1$ in
Planck units, as in the case of homogeneous cosmology. For the
inhomogeneous shell, we need to redefine the ``Planck time''. From
dimensional arguments, quantum gravitational effects become important
when the radius of curvature becomes of the order of
$l_p$~\cite{hbook}.  Therefore, we define that in an inhomogeneous
collapse, the ``Planck time'' is the hypersurface ${\cal R}=1$, where
$\cal R$ is the scalar curvature and its value is in Planck
units. This generalized ``Planck time'' is consistent with the Planck
time in homogeneous cases. For example, the scalar curvature of
metric~\meq{frf} is $1.33$ at the Planck time and conversely, ${\cal
  R}=1$ corresponds to $T=1.15$.  Solving ${\cal R}=1$ from the metric
of the shell~\meq{thm}, we find
\begin{equation} 
T= T_{Pl}(R)=\frac{1}{3}\Big(6BR-\sqrt 3 \sqrt{4+3B^2 R^2}\Big)\,.
\label{tpr}
\end{equation}
Now we have the observer-dependent ``Planck time'' $
T_{Pl}(R)$. Analogously, we require that the physical entropy $s(T,R)$
be equal to $1$ at $T=T_{Pl}(R)$. Therefore, \eq{ssr} yields
\begin{equation} 
s(T,R)=\frac{r'(T_{Pl}(R),R)r^2(T_{Pl}(R),R)}{r'(T,R)r^2(T,R)}\,. 
\label{sft} 
\end{equation}
Substituting \eqs{tpr} and~\meq{sft} into \eq{ssr}, we have the
following simple result
\begin{equation} 
S(R)=8\pi\,. 
\label{bsr}
\end{equation}
Thus the total entropy enclosed in the shell is
\begin{equation} 
S=8\pi(R_2-R_1)\,. 
\label{tey}
\end{equation}
Now we are ready to test the covariant entropy bound. First, we shall
choose the value of $M_0$. Then, we solve numerically for $R_2$ from
the equation of the radial null geodesic which starts at $G$ and ends
at the singularity.  Once we have obtained $R_2$, the total entropy
crossing $L$ can be obtained from \eq{tey} immediately (remember that
$R_1=M_0$). To check the bound, we define
\begin{equation} 
{\rm Ratio}\equiv \frac{4S}{A}\,,
\end{equation}
where $A=4\pi (2M_0)^2$ is the area of the light sheet. If ${\rm
  Ratio}\leq 1$, the bound is satisfied. Otherwise, it is
violated. Figure~\ref{ma:p1} shows that the bound holds for large
values of $M_0$ and the Ratio decreases with $B$. However,
figure~\ref{ma:p2} shows that violation occurs for $M_0<50$ and small
values of $B$. Since $M_0= 50$ is not near the Planck mass $m_{Pl}=1$,
it seems that the breakdown of classical relativity near the
singularity cannot account for the violation. In order to explain this
apparent violation, we consider the case of $B=0$, i.e., the flat
dust-filled universe solution~\meq{frf}, because figure~\ref{ma:p2}
indicates that the maximum violation occurs at $B=0$.

\FIGURE[t]{\epsfig{file=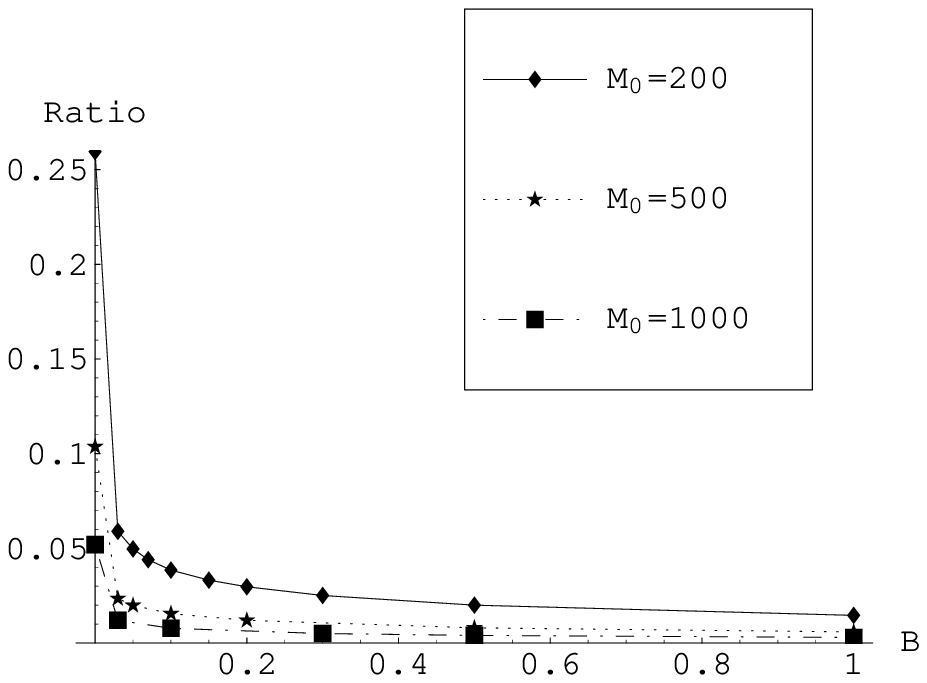, width=.6\textwidth}%
\caption{It is plotted the Ratio $\equiv {4\,S}/{A}$ as a function
  of the inhomogeneity parameter $B$, for three different large values
  of $M_0$. One sees that the entropy bound is satisfied for large
  values of $M_0$.\label{ma:p1}}}

\FIGURE[t]{\epsfig{file=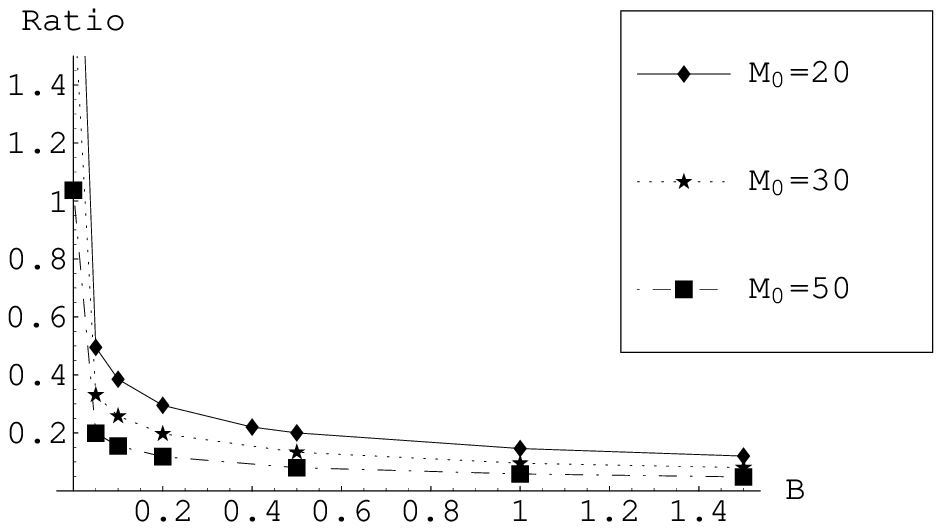, width=.65\textwidth}%
\caption{It is plotted the Ratio $\equiv {4\,S}/{A}$ as a function of
  the inhomogeneity parameter $B$, for three different small values of
  $M_0$.  For $M_0<50$, violations occur at small values of
  $B$.\label{ma:p2}}}

\FIGURE[t]{\begin{tabular}{p{.4725\textwidth}p{.4725\textwidth}}
{\epsfig{file=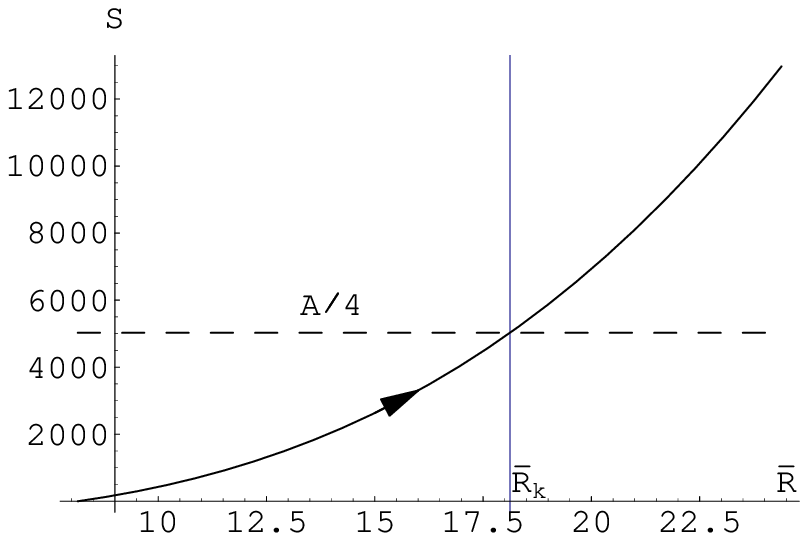, width=.4725\textwidth}}
&
{\epsfig{file=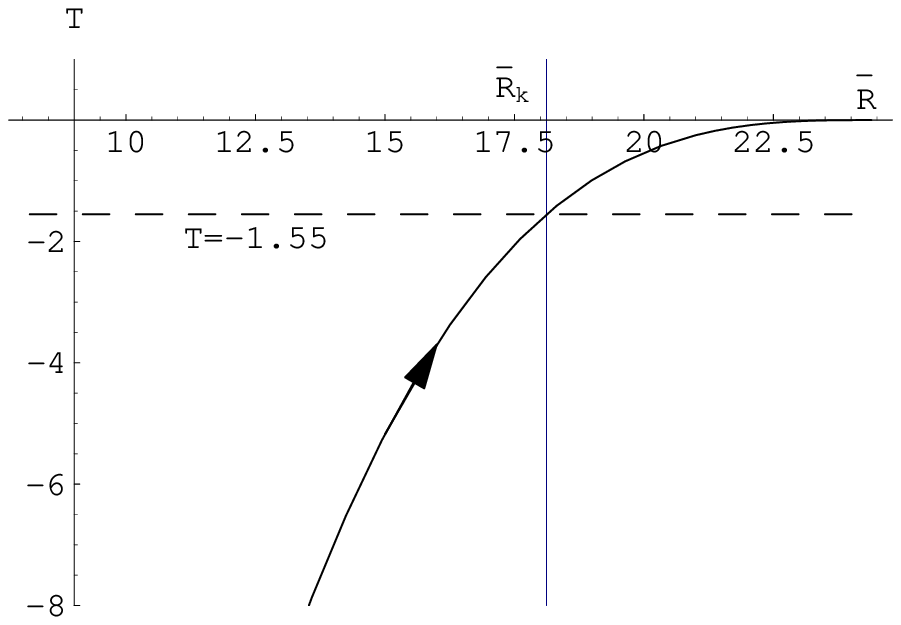, width=.4625\textwidth}}
\\
{\small $(a)$ The entropy crossing $L$ as a function of $\bar
  R$.}
&
{\small $(b)$ A plot of the light sheet $L$ in $(T,\bar R)$
  coordinates.}
	   \end{tabular}%
\caption{The entropy $S$ exceeds $A/4$ at $R>\bar R_k$. However, this
  region is within about a Planck time from the singularity, where
  a quantum description is necessary.\label{ent:fig}}}

The solution~\meq{frf} possesses a future crushing singularity at
$T=0$.  We choose a future-directed light sheet $L$ starting from a
2-sphere $B_0$ with coordinates $(T=-26.67, \bar R=8.143)$.  It is
easy to check, from \eq{rtr}, that this light sheet corresponds to
that in the Tolman-Bondi spacetime when $M_0=20$ (well below the
critical value $M_0=50$) and $B=0$. The 2-sphere is obviously not very
close to the singularity compared with the Planck scale. At $B_0$, the
scalar curvature ${\cal R}\approx0.00187$, which also indicates that
classical description is sufficient in the neighborhood of
$B_0$. However, as we shall now show, the violation of the bound is
due to the entropy crossing $L$ within about one Planck time from the
singularity, where quantum description is necessary.  Indeed,
figure~\ref{ent:fig}$a$ illustrates the entropy distribution along
$L$. The entropy saturates the $A/4$ bound at $R=\bar R_k\approx
18.12$. If we continue to count the entropy after $\bar R_k$ all the
way to the singularity, the bound will be exceeded by a wide
margin. However, figure~\ref{ent:fig}$b$ shows that $R=\bar R_k$ on
the light-sheet corresponds to $T=-1.55$, which is around the Planck
time. Since the portion of the shell crossing $L$ around this moment
is in the Planck regime where quantum gravity description is required,
it is reasonable to exclude the classical entropic reasonings in this
region. As the entropy crossing the light-sheet within about a Planck
length from the singularity should be excluded, the entropy bound
still holds.

\section{Discussions and conclusions}

\looseness=1 We have constructed an exact solution which corresponds to a
Tolman-Bondi dust shell collapsing into a Schwarzschild black hole
with mass $M_0$. The shell just crosses entirely the light-sheet $L$
which is terminated at the singularity. In the spirit of Bousso's
assumptions, the entropy in a comoving volume of the shell is assumed
to be a constant. We also set a future boundary condition, in contrast
to Bousso's past boundary condition, to specify the entropy. It is
always more difficult to argue on plausible final boundary conditions
than on initial ones. However, time symmetry (an issue far from being
settled) can be invoked to impose time symmetric boundary conditions.
Our numerical results show that for $M_0\gtrsim 50$, the entropy of
the shell does not exceed one quarter of the area of the black hole,
while for $M_0\lesssim 50$, the entropy exceeds one quarter of the
area. The later case appears to violate the covariant entropy bound.
However, this violation can be interpreted as over-counting the
entropy residing in the \pagebreak[3] Planck regime (very close to the singularity).
This part of entropy should not be included because classical
relativity breaks down in that region. Therefore, we conclude that there
is no violation on the covariant entropy bound during the gravitational
collapse.

\acknowledgments

This work was partially funded by Funda\c c\~ao para a Ci\^encia e
Tecnologia (FCT) --- Portugal through project SAPIENS 36280/99. SG
acknowledges financial support from FCT grant SFRH/BPD/10078/2002.
JPSL thanks Observat\'orio Nacional do Rio de Janeiro for hospitality.

\end{document}